# A Modular Design of Continuously Tunable Full Color Plasmonic Pixels with Broken Rotational Symmetry

*Rui Feng[1, †], Hao Wang[2, †], Yongyin Cao[1], Ray J. H. Ng[3], You Sin Tan[2], Yanxia Zhang[1], Fangkui Sun[1], Cheng-Wei Qiu[4, \*], Joel K. W. Yang[2,5, \*], Weiqiang Ding[1, \*]*

[1]Institute of Advanced Photonics, School of Physics, Harbin Institute of Technology, Harbin, 150080, China

[2]Engineering Product Development Pillar, Singapore University of Technology and Design, 8 Somapah Road, Singapore, 487372, Singapore

[3]Institute of High Performance Computing, A*STAR (Agency for Science, Technology and Research), 1 Fusionopolis Way, #16-16 Connexis, Singapore, 138632, Singapore

[4]Department of Electrical and Computer Engineering, National University of Singapore, 4 Engineering Drive 3, Singapore, 117583, Singapore

[5]Institute of Materials Research and Engineering, A*STAR (Agency for Science, Technology and Research), 2 Fusionopolis Way, #08-03 Innovis, Singapore, 138634, Singapore

**Abstract:** Color tuning is a fascinating and indispensable property in applications such as advanced display, active camouflaging and information encryption. Thus far, a variety of reconfigurable approaches have been implemented to achieve color change. However, it is still a challenge to enable a continuous color tuning over the entire hue range in a simple, stable and rapid manner without changes in configuration and material properties. Here, we demonstrate an all-optical continuously tunable plasmonic pixel scheme via a modular design approach to realize polarization-controlled full color tuning by breaking the intrinsic symmetry of the unit cell layout. The polarization-controlled full color tunable plasmonic pixels consist of three different types of color modules oriented at an angle of 60° with respect to each other, corresponding to three subtractive primary colors. Without changing the structural properties or surrounding environment, the structural colors can be continuously and precisely tuned across all hues by illuminating linearly polarized light with different polarization directions. Meanwhile, the plasmonic pixels can be flexibly customized for various color tuning processes, such as different initial output colors and color tuning sequences, through the appropriate choice of component modules and the elaborate design of module layouts. Furthermore, we extend the color tuning to achromatic colors, white or black, with the utilization of a single module or the introduction of a black module. The proposed polarization-controlled full color tunable plasmonic pixels hold considerable potential to function as next-generation color pixels integrated with liquid-crystal polarizers.

Structural colors generated from the resonant interactions between incident visible light and sophisticated artificial nanostructures have recently become one of the most interesting scientific and engineering topics in the field of optics, and have applications in high-resolution color prints, high-efficiency color filters, high-contrast color displays, anti-counterfeiting tags, and information encryption[1-8]. In contrast to conventional pigmentary colors produced by bulk absorption, structural colors provide many benefits such as high spatial resolution, good stability, material simplicity, cost effectiveness, non-toxicity, and recyclability[9-10]. An assortment of vivid and vibrant structural colors covering the entire visible spectrum can be engineered by tailoring the spectral responses, which is strongly dependent on the size and shape of the constituent nanostructures, as well as their arrangement[11-14]. However, this geometric dependence of structural colors makes it difficult to change the static colors once the nanostructures are fabricated with given sets of dimensions which severely restricts their potential applications in many areas. It is thus an outstanding problem to finely tune structural color across a full range of hues, while maintaining high spatial resolution, real-time response and high durability[15-16].

With the development of manufacturing technologies and advanced materials, research on reconfigurable structural colors have progressed dramatically in recent years[17-18]. The most straightforward method for the realization of color tuning is to vary the geometric properties of the nanostructures, such as size, shape or period. For instance, laser-induced heating has been explored for ink-free, high-speed and sub-diffraction resolution color printing[19-21]. In this case, the plasmonic heating utilizes the photothermal effect of plasmonic resonance energy absorption to locally melt and reshape each nanostructure[20]. Likewise, dielectric heating arising from dipole resonance energy absorption in high-index dielectric germanium (Ge) causes a similar heating effect[21]. This laser post-processing technique will irreversibly print various new colors over nanostructured metasurfaces at an unprecedented resolution. Conversely, the reversible electrodeposition scheme was demonstrated to manipulate the colors for active artificial camouflage[22]. The bimetallic nanodot arrays were altered through electrodepositing and stripping Ag shells on top of Au nanodomes. Thus, the reversible color plasmonic display can be achieved by electrochemically controlling the deposition and stripping duration. In addition to the morphological reconfiguration of the nanostructures, the modification of the nanostructural arrangement by stretching the underlying elastomeric substrate was also shown color tuning and multiple image switching[23]. Alternatively, the color tuning can be realized by changing the dielectric properties of the constituent nanostructures or the surrounding environment[24-26]. In particular, through controlled hydrogenation and dehydrogenation, constituent magnesium nanoparticles in a nanostructure pixel undergo reversible transitions between metallic and dielectric states, which can be exploited for dynamic color display and information encryption[27-28]. Similarly, by applying an appropriate voltage, the dielectric properties of the surrounding environment could be changed significantly due to reversible switching between the oxidized and reduced states of the electrochromic materials[29], changes in the molecular orientation of birefringent liquid crystal[30-32], and voltage-driven hydrogenation and dehydrogenation cycles in a solid-state device[33], which supports large color changes with potential for fast switching[34]. Furthermore, the refractive index of the surrounding media can also be directly changed by injecting different chemical solutions into a polymeric microfluidic channel which covers the metasurfaces[35].

Although previous methods of varying the geometric properties and changing the dielectric properties introduced above can effectively tune the colors, few solutions have achieved full color tuning, while some exhibit switching only between color states[36-37]. Moreover, most solutions require the employment of complex packaging devices, such as electrochemical cells, gas cells or microfluidic cells, which result in increased fabrication cost and complexity[27-29, 35]. Furthermore, some techniques also suffer from a relatively long switching times, poor reliability and durability, or are limited by a one-time color shift, e.g. in the case of laser post-processing[20-21]. Recently, the modular design approach has been widely applied in photonics, which is a very practical and flexible strategy[38-39]. For effective modular design, a system is subdivided into smaller parts called modules which can be independently created, tailored or replaced between different systems to drive different functionalities. In this letter, we propose a plasmonic pixel whose color can be continuously tuned over the entire hue range by adopting the modular design approach in which independently tailored plasmonic nanoantennas serve as color modules and are combined into a composite unit cell. In contrast to the prior dual color pixels based on the polarization-dependent cross-shaped nanoantennas or nanoapertures that have limitation in terms of achieving full-color tuning and are limited to switching between a limited range of hues [40-43], we construct an asymmetric nanostructure by combining three different types of color modules corresponding to three subtractive primary colors. Without the aid of extra active materials or external stimuli[44], the proposed polarization-controlled full color tunable plasmonic pixels can generate a full range of hues by a continuous change in the polarization angle of the incident light. Furthermore, by taking full advantage of the modular design approach, we can flexibly customize the color tuning process to change the initial output colors and the color tuning sequences, and even to switch between two primary colors or between chromatic and achromatic colors.

**RESULTS AND DISCUSSION**

The perceived colors, which can be reproduced by the mixture of three primary colors, are located within a triangle with vertices corresponding to these three primary colors on the chromaticity diagram. Similarly, in order to construct the full color tunable plasmonic pixels, three different primary color modules, which are polarization dependent plasmonic nanoantennas with different lengths and orientations, are combined in a subwavelength unit cell. However, they are not excited together at the same polarization angle, as each primary color module only dominates at a specified polarization angle depending on the rotation angle of the plasmonic nanoantenna. The schematic diagram of the polarization-controlled full color tunable plasmonic pixel (FCTPP) is illustrated in Figure 1a. The rectangular aluminum and silica nanorods with three different lengths are employed to function as the color modules for yellow, magenta and cyan respectively which are determined by the absorption resonance wavelengths in the subtractive color model. Noticeably, the absorption cross-section of the nanoantenna for the yellow module is smaller than those for the magenta and cyan modules due to its relatively small size[45]. To compensate for the low absorption efficiency and the narrow bandwidth at short wavelengths for the yellow module, two identical nanoantennas that function as the yellow module are arranged along the diagonal quadrants of the unit cell. As a comparison, we also investigate the unit cell with only one nanoantenna for the yellow module, which has a smaller color gamut (Supplementary Figure S1). The lengths of the yellow, magenta, and cyan modules are 95 nm, 120 nm, and 140 nm respectively, corresponding to plasmonic resonances at different wavelengths across the visible spectrum. Most importantly, the three different types of color modules are rotated by an angle of 60° with each other to form an asymmetric nanostructure (the rotation angles of the color modules are 0° for the yellow module, 60° for the magenta module and 120° for the cyan module), making the resulting nanostructures sensitive to the linear polarization angle of the incident light. Considering that a 180° rotation of the polarization direction is divided up into three states, the modules are arranged in a 60° interval to minimize the spectral mixing that arise from the partial excitation of at least double color modules. For instance, with the polarization of incident light aligned to the long axis of one of the modules, the non-zero decomposition of this polarization along the long axis of the other two modules would still lead to partial excitation of their longitudinal resonance modes. These four plasmonic nanoantennas are densely packed at a fixed center-to-center distance of half the unit-cell period. The period of the composite unit cell is set to be 360 nm, which suppresses grating effects and ensures that the print resolution is close to the diffraction limit for a high degree of color homogeneity. The thickness of the aluminum and silica nanorods are both 30 nm, and the widths of the strips are 50 nm for all nanoantennas. Here the polarization angle $\varphi$ is defined as the angle between the direction of the electric field and the $x$-axis of the coordinate system, as shown in the inset of Figure 1a.

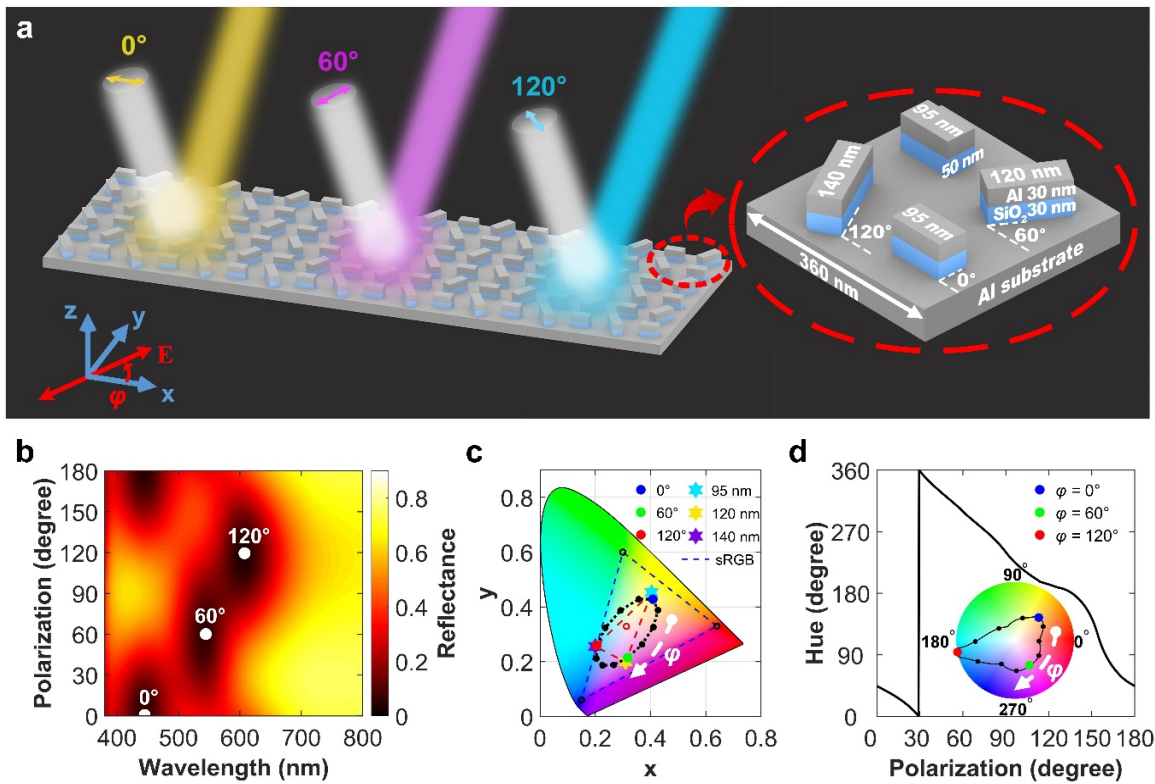

**Figure 1. (a)** Schematic of polarization-controlled full color tunable plasmonic pixels for generating a complete hue of colors when illuminated with different polarization of the incident

light. The proposed plasmonic pixels are made of asymmetric nanostructures comprising three different types of color modules (two 95 nm antennas repeated) oriented at 60° rotations relative to each other. The lengths of the three color modules are 95 nm, 120 nm, and 140 nm, respectively, and the widths are 50 nm for all nanoantennas **(b)** Simulated reflectance spectra of the full color tunable plasmonic pixels as a function of polarization angle at normal incidence. Three reflection dips are indicated by the white dots at the polarization angles of 0°, 60°, and 120°. **(c)** Cyclic path of color coordinates for different illuminating polarization angles increasing from 0° to 180° on the CIE 1931 chromaticity diagram. The color coordinates of the pixels made up of only one type of color module with lengths of 95 nm, 120 nm and 140 nm (stars) lie near those of polarization-controlled full color tunable plasmonic pixels at the polarization angles of 0°, 60° and 120° (points). Here, the marks of the different colors are used to distinguish the close chromaticity coordinates. Dashed white arrow indicates the color tuning trajectory with increasing polarization angle of the incident light. **(d)** Achievable hue variation of the proposed plasmonic pixels with varying the incident polarization angle, demonstrating the continuous color tuning across all hues. The inset shows the calculated colors for different polarization angles in HSV color space.

The proposed polarization-controlled FCTPPs can reflect different colors when illuminated by linearly polarized light with the polarization angle changing from $\varphi = 0°$ to 180°. As shown in Figure 1b, three reflection dips are located at the polarization angle of $\varphi = 0°$ near 440 nm, $\varphi = 60°$ near 540 nm, and $\varphi = 120°$ near 610 nm, respectively. These three reflection dips correspond to three subtractive primary colors and the polarization dependent reflectance spectra will result in different colors based on subtractive color mixing. To understand the physical mechanism of these three reflection dips, we present the electric and magnetic field distributions at resonances for three different polarization angles, which clearly demonstrate relatively independent excitation of gap-surface plasmons within the corresponding plasmonic nanoantenna[46] (Supplementary Figure S2). Furthermore, each plasmonic nanoantenna can confine the light strongly within the spacer region, which limits plasmonic coupling with adjacent nanoantennas (Supplementary Figure S3). Hence, the origin of this polarization-controlled full color tuning lies in the intrinsic polarization-dependent excitation of gap surface plasmons in individual structures. To evaluate the performance of the FCTPP, the chromaticity coordinates of the reflectance spectra for different polarization angles are calculated and plotted as black dots on the CIE 1931 chromaticity diagram, as shown in Figure 1c. The CIE "white light" illuminant D65 is used for conversion to the chromaticity coordinates. It is clearly seen that the chromaticity coordinates gradually shift from yellow towards magenta, then to cyan, and finally back to yellow again when the incident polarization angle increases from 0° to 180°, as indicated by the dashed white arrow. The polarization-controlled full color tuning is represented by the black dotted ellipse around the white point on the chromaticity diagram, which is attractive in comparison with the static color pixels or dual color pixels that generate a limited range of colors represented by a single point or a short line on the chromaticity diagram[40, 47], showing that the FCTPP can be rapidly and precisely modulated to generate a wide variety of colors over the entire hue range. We also show three stars on the chromaticity diagram that represent the chromaticity coordinates if the entire unit cell is made up of only one type of color module. The close chromaticity coordinates of the marks indicate that the partial excitations of the other two color modules caused by the non-zero decomposition of incident polarized light will not strongly influence the main resonant dips in the reflectance spectra, but only lead to low sideband reflectance. Therefore, the partial excitations of color modules in the asymmetric nanostructure will not result in a significant color change. Interestingly, the elliptical color tuning trajectory with increasing polarization angle lie outside the triangle that connect these three stars, which is caused by the mixed spectra of the resonance dips for the asymmetry structure[48] (Supplementary Figure S4). In addition, the FCTPP allows for continuous color tuning across all hues by finely controlling the polarization angle of the incident light, which is impossible for the methods of varying the geometric properties or changing the dielectric properties[22, 27], as shown in Figure1d. Owing to the principle of the subtractive color model, the color saturation of the proposed FCTPP is smaller than that of the dielectric color pixels based on the additive color model, but the gamut for polarization-controlled full color tuning is comparable in area to that for the conventional static color change by varying the nanoantenna length (Supplementary Figure S5).

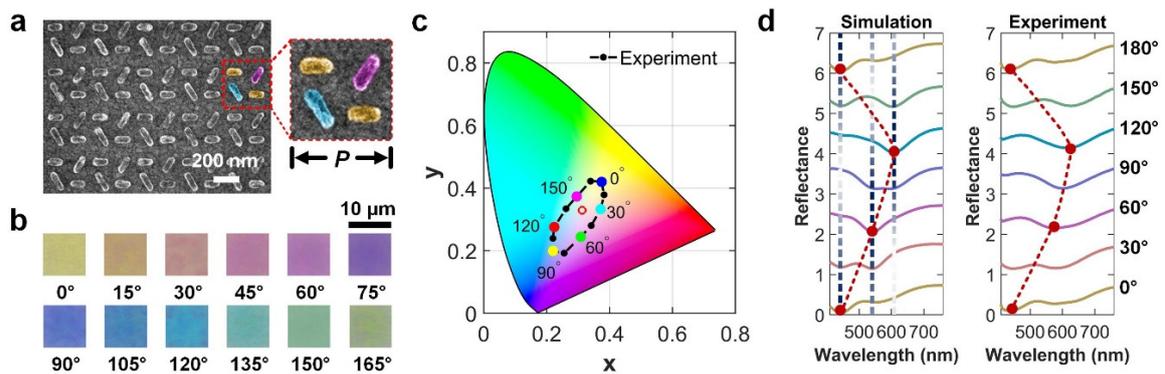

**Figure 2. Optical characterization of the polarization-controlled full color tunable plasmonic pixels. (a)** SEM images of the fabricated plasmonic pixels consisting of aluminum and silica ($SiO_2$). The enlarged SEM image of a single unit cell shows three different types of color modules with different lengths and orientations. **(b)** Bright-field optical micrographs of the fabricated color palettes labeled with the corresponding polarization angles of the incident light. Each square is $10\times10$ $\mu m^2$. **(c)** Chromaticity coordinates of the fabricated color palettes for different polarization angles, which demonstrate the full color outputs. **(d)** Simulated and experimental reflectance spectra of the plasmonic pixels for different polarization angles.

To verify the design, the asymmetric nanostructures were fabricated using high-resolution electron beam lithography and electron beam evaporation (details in Methods). The scanning electron microscope (SEM) images of the fabricated plasmonic pixels are shown in Figure 2a. The four nanoantennas of different rotation angles and lengths are well defined, with minimal imperfections such as rounded corners and position deviations. Figure 2b presents the optical microscope images of the FCTPPs for different polarization angles ranging from 0° to 180° in steps of 15°. The measured reflection color palettes show a series of colors covering a wide gamut. Notably, three subtractive primary colors yellow, magenta, and cyan are obtained at the polarization angles of 0°, 60° and 120°, respectively. Figure 2c shows the corresponding chromaticity coordinates of fabricated color palettes on the chromaticity diagram, which experimentally validates that a full range of colors can be obtained at different polarization directions with a single plasmonic device. Figure 2d shows the simulated and measured reflectance spectra of the FCTPPs for different polarization angles increasing from 0° to 180°. The experimental reflectance spectra are found to be in good agreement with the simulation results. In contrast to the gradual redshift of the resonance wavelength caused by changing the structural properties or surrounding environment, the reflection dips produced by the asymmetric nanostructure drastically change among three resonance wavelengths by varying the incident polarization angle, as indicated by the red dots. The three reflection dips are most prominent at the polarization angles of 0°, 60° and 120°, respectively, which depend on the rotation angles of the color modules. Apparently, the polarization dependent reflectance spectra of the FCTPPs are related to a linear combination of the polarization dependent absorption of three single color modules, which verify the strong confinement property of gap surface plasmons and the Malus's

Law of polarization-controlled color tuning (Supplementary Figure. S6). Furthermore, we also investigate the incident angular tolerance of the polarization-controlled FCTPPs (Supplementary Figure. S7). The reflectance spectra of the FCTPPs for different polarization angles are approximately unaffected when the incident angle is up to 40°. Thus, the polarization-controlled full color tuning originated from the asymmetric nanostructure is shown to be maintained at oblique incidence, and these contribute to the expanded viewing angle of our system. [25, 47]

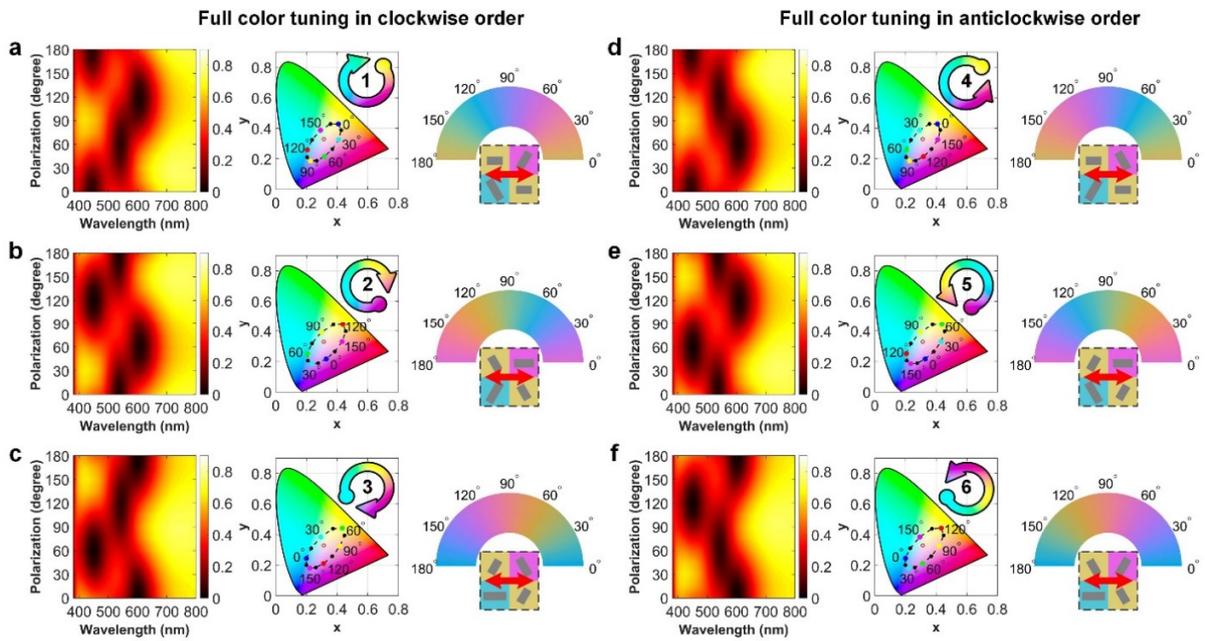

**Figure 3. Simulations of various color tuning sequences achieved with different module layouts. (a)-(c)** Polarization-controlled full color tuning with the initial color of yellow, magenta and cyan in clockwise order on the chromaticity diagram. **(d)-(f)** Polarization-controlled full color tuning with the initial color of yellow, magenta and cyan in anticlockwise order on the chromaticity diagram. The initial color is determined by the color module oriented along the polarization angle 0°, while the color tuning sequence is related to the rotation angles of the other two color modules. The insets show the corresponding module layouts for various color tuning processes. The red arrow represents the initial polarization angle of the incident light. The colors of the point are blue,

cyan, green, yellow, red and magenta for the polarization angle of 0°, 30°, 60°, 90°, 120°, and 150°, respectively.

On the basis of flexible and configurable properties of the modular design approach adopted for the proposed polarization-controlled FCTPPs, we can flexibly customize the color tuning process by using various module layouts. As a demonstration, we present the FCTPP with different initial output colors and color tuning sequences. Considering that a 180° rotation of the polarization direction is divided up into three states, each color module can be oriented along the *x*-axis, meanwhile the other two color modules are arranged for the remaining states. Thus, the number of possible module layouts is the factorial of 3, i.e., 3! = 6 for three different types of color modules oriented along the rotation angles of 0°, 60° and 120°. It is firstly observed that the initial output colors, which refer to the colors for the polarization angle of 0°, are quite different for various module layouts, as shown by the blue dots on the chromaticity diagram of Figure 3a-c. The initial output colors are yellow, magenta and cyan when the respective color modules align with the *x*-axis of the coordinate system, because the resonant absorption by the corresponding color modules are dominant at the polarization angle of 0° (as shown in the inset). In fact, the initial output color can be set to any hue on the chromaticity diagram, we also present FCTPP with the initial color of red and green on the chromaticity diagram. (Supplementary Figure S8). Thereafter the chromaticity coordinates of these three full-color tunable plasmonic pixels are tuned in clockwise order on the chromaticity diagram as the incident polarization angle increases. Nonetheless, as shown in Figure 3d-f, the color tuning sequence can also be reversed to anticlockwise on the chromaticity diagram by swapping the rotation angles of the color modules. Taking Figure 3d as an example, the color tuning trajectory evolves from yellow to cyan and then to magenta with increasing polarization angle of the incident light, which is in the opposite order to that of the plasmonic pixels in Figure3a. In general, the color tuning sequences depend on the rotation angles of each color module relative to the *x*-axis. In addition, the color tuning range can also be limited to a tuning between two primary colors according to the choice and combination of two distinct types of color modules from the set of yellow, magenta and cyan modules. Notably, the three different module layouts can achieve color tuning from one to another by changing the polarization angle of the incident light, similar to the dual-color pixels (Supplementary Figure S9). Consequently, various color tuning processes can be generated through the appropriate choice of component module and the elaborate design of module layouts. The color tuning process become customizable and are no longer a simple redshift or blueshift change.

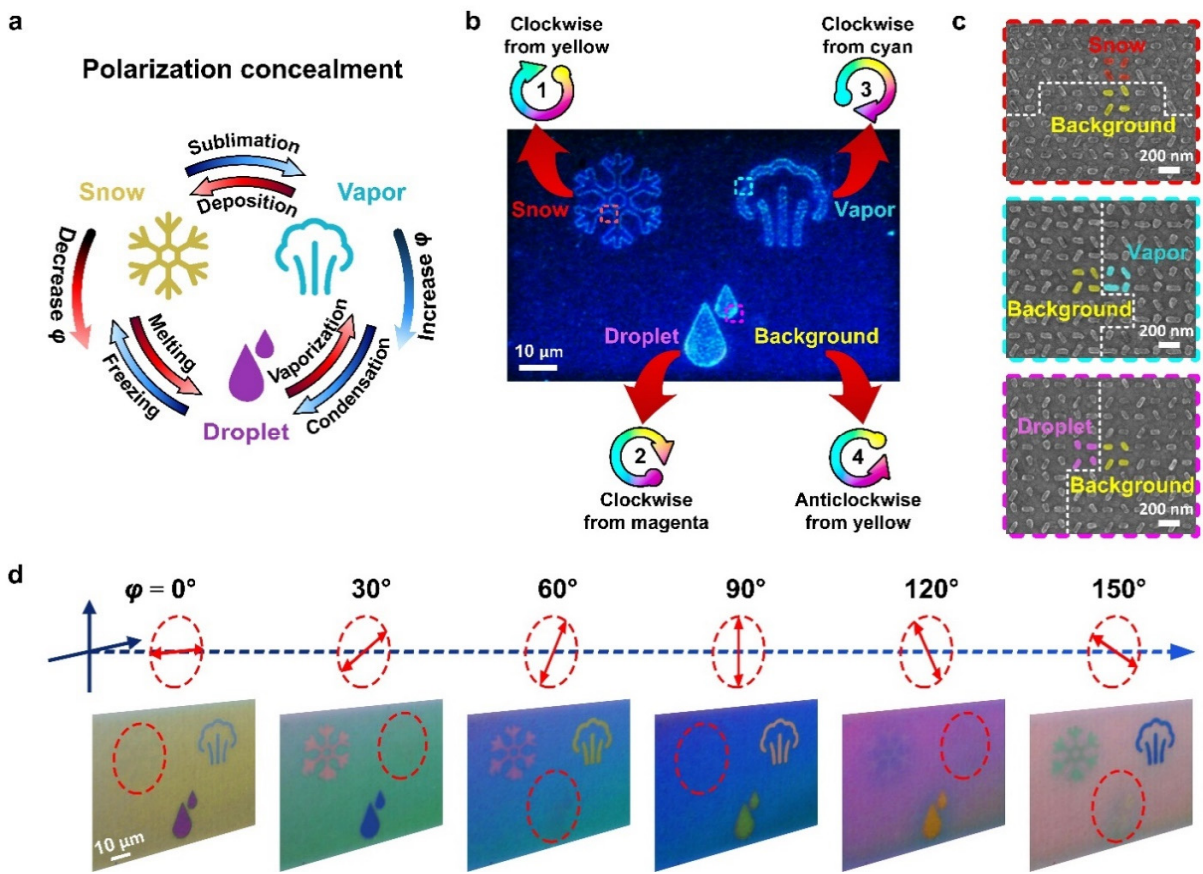

**Figure 4. Demonstration of the polarization concealment. (a)** Schematic illustration of the polarization concealment. Three patterns of snow, vapor, and droplet, which represent three states of water, can be concealed in turn with changing the polarization angle of the incident light. In this instance, the polarization angle is 0°, so the snow pattern is concealed, while the remaining two patterns that represent the liquid and gaseous phases of water are visible. **(b)** Dark-field optical micrograph of the polarization concealment microprint. The background and three patterns are labeled with the corresponding module layouts and their calculated color palettes. **(c)** The SEM images of the borders between different module layouts for the patterns and the background. **(d)** Optical micrographs of the fabricated microprint at different polarization directions. The three clockwise-tuned patterns will meet the anticlockwise-tuned background twice on the chromaticity diagram within a 180° rotation of the polarization direction, i.e., 0° and 90° for the snow, 30° and 120° for the vapor and 60° and 150° for the droplet. Thus, the patterns of snow, vapor and droplet are each concealed twice for different polarization angles ranging from 0° to 180°.

The customizable characteristics of the color tuning process make such polarization-controlled FCTPPs highly suitable for information encryption. The target images are encoded by various module layouts for the background and the emerging patterns, then the patterns can be concealed or revealed by controlling the polarization angle of the incident light. Figure 4a presents the schematic illustration of the polarization concealment. Three patterns of snow, vapor and droplet, which represent three states of water (solid, gaseous, and liquid phases), can be concealed in turn by using incident light with the appropriate polarization angle. Along the clockwise direction, the phase transitions are: vapor → droplet (condensation), droplet → snow (freezing), and snow → vapor (sublimation), while changing the polarization rotation direction, i.e., along the anticlockwise direction, the phase transitions are: vapor → snow (deposition), snow → droplet (melting), and droplet → vapor (vaporization). To demonstrate this capability, we fabricated the polarization concealment microprint using different plasmonic pixels with the module layouts derived in Figure 3a-d. Three FCTPPs which are tuned clockwise on the chromaticity diagram with different initial output colors of yellow, magenta and cyan are used for the three patterns, whereas the FCTPP placed in the background are tuned anticlockwise on the chromaticity diagram with an initial output color of yellow. The dark-field microscopic image reveals the outlines of the three patterns in the background with the corresponding color tuning sequences, as shown in Figure 4b. The SEM images for the boundaries of two different module layouts are shown in Figure 4c. From Figure 4d, we can see the polarization direction for optical micrographs of the polarization concealment microprint advance anticlockwise with increasing polarization angle. Evidently, the color of the three patterns take turns to be consistent with the background with changing the polarization angle of the incident light, exhibiting the corresponding phase transitions in Figure 4a. The designed microprint demonstrates the polarization concealment that conceals specific information inside the background when the microprint is illuminated with light of a specific polarization direction. In addition, we also present the optical micrographs of the microprint under illumination by natural light. It is observed that the three patterns can be well concealed in the background (Supplementary Figure S10).

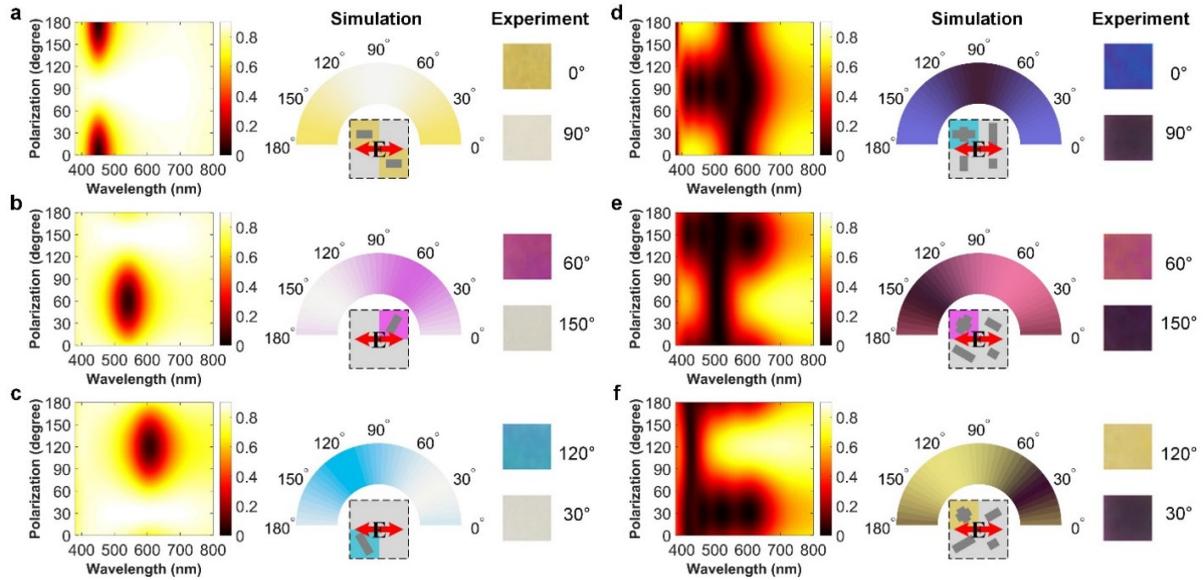

**Figure 5. Polarization-controlled color switching between achromatic colors and three subtractive primary colors**. **(a)-(c)** Simulated reflectance spectra and evolutions of the color switching between white and three subtractive primary colors of yellow, magenta and cyan. **(d)-(f)** Simulated reflectance spectra and evolutions of the color switching between black and three subtractive primary colors of cyan, magenta and yellow. The red arrow represents the initial polarization angle of the incident light. The experimental optical micrographs show the achromatic colors and three subtractive primary colors at the specified polarization angles. The insets present the corresponding module layouts for various color switching modes.

Remarkably, the polarization-controlled FCTPPs afforded by the modular design approach enable color tuning not only among different color hues, but also between chromatic and achromatic colors, i.e., white or black. Following the modular design approach, we first investigate the relatively simple color switching between white and three subtractive primary colors, as shown in Figure 5a-c. Because the color modules for three subtractive primary colors can function independently, only one type of polarization dependent color module is employed in the unit cell for the color switching. The corresponding module layouts are shown in the inset with the labels indicating the polarization angle of the incident light. The lengths and rotation angles of the plasmonic nanoantennas for the three module layouts are the same as those of the plasmonic pixels aforementioned in Figure 1 (95 nm and 0° for the yellow module, 120 nm and 60° for the magenta module, and 140 nm and 120° for the cyan module). As shown in the simulated reflectance spectra and color palettes of Figure 5a-c, the three subtractive primary colors are produced at the specific polarization angles of $\varphi = 0°$, 60°, and 120°, respectively, which are consistent with the rotation angles of the plasmonic nanoantennas. Whereas bright white color can be obtained at the orthogonal angles. The experimentally obtained optical micrographs at the specified polarization angles show good agreement with the simulation results (The experimental reflectance spectra are shown in Supplementary Figure S11). On the other hand, the color switching between black and three subtractive primary colors is much more difficult, as shown in Figure 5d-f. The generation of black requires strong light absorption over a broad spectral range, thus four different color modules with the same rotation angles are utilized as a black module to cover the visible region. The lengths of four different nanoantennas in the four quadrants responsible for the black modules are 80 nm, 95 nm, 120 nm, and 140 nm, respectively. The nanoantennas of 140 nm and 80 nm are arranged on one side to avoid excessive near field plasmonic coupling. Whereafter, another primary color module is placed in the orthogonal direction to form a cross-shaped nanoantenna for the color switching, as shown in the inset. The lengths and rotation angles of the orthogonal arms are 140 nm and 0° for the cyan module, 120 nm and 60° for the magenta module, and 95 nm and 120° for the yellow module, respectively. Similarly, the three subtractive primary colors appear at the specific polarization angles of $\varphi = 0°$, 60° and 120°, respectively, while a distinct black color can be observed at the orthogonal angle. Again, the experimental optical micrographs at the specified polarization angles exhibit a close agreement with the calculated colors. Notably, the quality of black color can be further improved through optimizing the nanoantenna lengths or employing high-loss dielectric gap layers to increase the light absorption. For various color switching modes from the subtractive primary colors to achromatic colors, the plasmonic pixels will undergo a gradual color fading with changing the incident polarization angle, and slowly turn to white or black at the orthogonal angle with the chromatic color vanishing. The initial states of the color switching can be customized to be either subtractive primary colors or achromatic colors by selecting the appropriate rotation angles of the color modules.

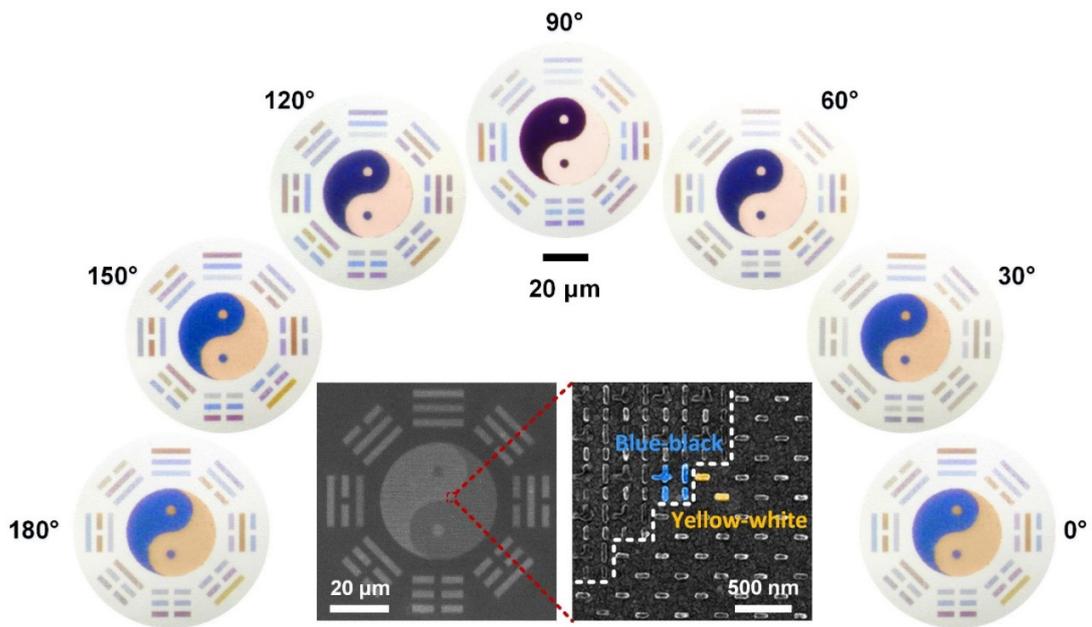

**Figure 6. Demonstration of color display with achromatic colors.** Optical micrographs of the fabricated Bagua diagram show the various processes of color change by changing the polarization angle of the incident light. The outer trigram patterns and the inner Taiji pattern consist of full color tunable plasmonic pixels and chromatic-achromatic color switchable pixels, respectively. The SEM images of the fabricated Bagua diagrams with an enlarged view of the selected region illustrating the border between two module layouts.

To demonstrate the feasibility of the various processes of color change, we design a "Bagua diagram" using different polarization-controlled FCTPPs. Figure 6 presents the optical micrographs of the Bagua diagram when illuminated with different polarization directions. The outer trigram patterns consist of FCTPPs with different initial output colors of yellow, magenta and cyan, respectively. The inner Taiji patterns comprise two color switchable pixels that can switch between primary colors and achromatic colors, namely blue-black color switchable pixels for the left spiral and yellow-white color switchable pixels for the right spiral. The overall SEM image and enlarged SEM image of the border between two regions are shown in the inset of Figure 6. It is noted that the colors of the trigram patterns significantly change with increasing polarization angle of the incident light. Meanwhile the colors of the Taiji pattern would first fade to achromatic colors as the incident polarization angle increases to 90°, and then turn back to the original states when the incident polarization angle becomes 180°. Obviously, the polarization-controlled FCTPPs with different module layouts accurately reproduce various color changes.

**CONCLUSION**

In conclusion, we have presented polarization-controlled full color tunable plasmonic pixels based on the combination of three different types of color module in a subwavelength unit cell. By virtue of the strong confinement and polarization dependent properties of the gap-surface plasmons, the asymmetric plasmonic pixels enable polarization-controlled full color tuning by continuously varying the polarization angle of the incident light. These polarization-controlled color responses can yield an elliptical trajectory covering all hues on the chromaticity diagram. In addition, the color tuning process, such as initial output colors and color tuning sequences, can be flexibly customized through the appropriate choice of component module and the elaborate design of module layouts. Meanwhile, the color tuning between two primary colors or even switching between chromatic and achromatic colors can be realized by means of suitable functional color modules and specific module layouts. We believe that our scheme could provide a promising platform for applications in advanced display technology, optical security devices and optical data storage.

**METHODS**

**Fabrication.** A 100 nm thick film of aluminum was first evaporated onto a bare silicon substrate using an electron beam evaporator (Labline, Kurt J. Lesker Co.). The working pressure was $1.2 \times 10^{-6}$ Torr and the deposition rate was 1.5 Å/s. Then the positive-tone electron-beam resist PMMA (950K A4, MicroChem Corp.) was spin-coated onto the Al film at 5000 rpm for 60 s and then baked at 180 °C for 90 s. Electron beam lithography was performed using the eLine Plus, Raith GmbH at 30 kV acceleration voltage and 360 pA beam current with a $100 \times 100$ $\mu m^2$ write field. Proximity effect correction was performed before exposure. In the next step, the exposed samples were developed in 1:3 methyl isobutyl ketone/isopropanol (MIBK/IPA) solution at 0 °C for 45 s, followed by rinsed in IPA for 5 s and blow-dried using nitrogen gas. Afterwards, the silica layer and a second Al layer were sequentially deposited with the same electron beam evaporator under the same conditions, Finally, a lift-off process was performed to remove the unexposed resist by soaking the samples in acetone at 60 °C, followed by rinsing with IPA for 10 s and dried with nitrogen.

**Characterization.** Scanning electron microscopy was performed with JEOL-JSM-7600F at 5 kV acceleration voltage and Raith eLine Plus systems. at 10 kV acceleration voltage. The reflectance spectra and optical micrographs were taken via a Nikon Eclipse LV100ND optical microscope equipped with a Nikon DS-Ri2 camera and a CRAIC 508 PV microspectrophotometer. A halogen lamp (LV-HL 50 W) was used to illuminate the samples in reflection mode. The reflectance spectra were normalized with an aluminum mirror. The experimental spectra, bright-field and dark-field micrographs were taken with a Nikon 20×, NA = 0.45 objective lens.

**Simulation.** The reflectance spectra, and electric and magnetic fields were obtained with a commercial Finite-Difference Time-Domain software (Lumerical, Canada). For periodic array simulations, a standard plane wave source was employed with periodic boundary conditions along the $x$- and $y$-axes. Perfectly matched layers (PML's) were used for top and bottom boundaries, which are perpendicular to the propagation direction. The source spanned the wavelength range of 380 to 800 nm. The permittivity data of aluminum and silica were obtained from Palik.

## ASSOCIATED CONTENT

Supporting Information

Compensation for the yellow module, Spatial distributions of the fields, Comparison of color tuning trajectory, Comparison of color gamut with static color, The phenomenological expression, Incident angular tolerance, Initial color of red and green, Different color tuning ranges, Microprint under natural light, Reflectance spectra of achromatic color.

## AUTHOR INFORMATION

### Corresponding Authors

*E-mail: chengwei.qiu@nus.edu.sg

*E-mail: joel_yang@sutd.edu.sg

*E-mail: wqding@hit.edu.cn

**Author Contributions**

Rui Feng and Hao Wang contributed equally to this research.

**Notes**

The authors declare no conflict of interest.

**Acknowledgements**

This work was supported by National Natural Science Foundation of China (Grants No. 11704088 and No. 11874134). China Postdoctoral Science Foundation (Grant No.2017M621255), Heilongjiang Postdoctoral Fund (LBH-Z17071), Fundamental Research Funds for the Central University (Grant No. HIT.NSRIF.2019059). We thank the HPC Studio at Physics Department of Harbin Institute of Technology for access to computing resources through INSPUR-HPC@PHY.HIT. This work was also supported by National Research Foundation (NRF) Singapore, under its Competitive Research Programme NRF-CRP001-021, Singapore University of Technology and Design (SUTD) Digital Manufacturing and Design (DManD) Center (RGDM 1830303).

**Supplementary Information**

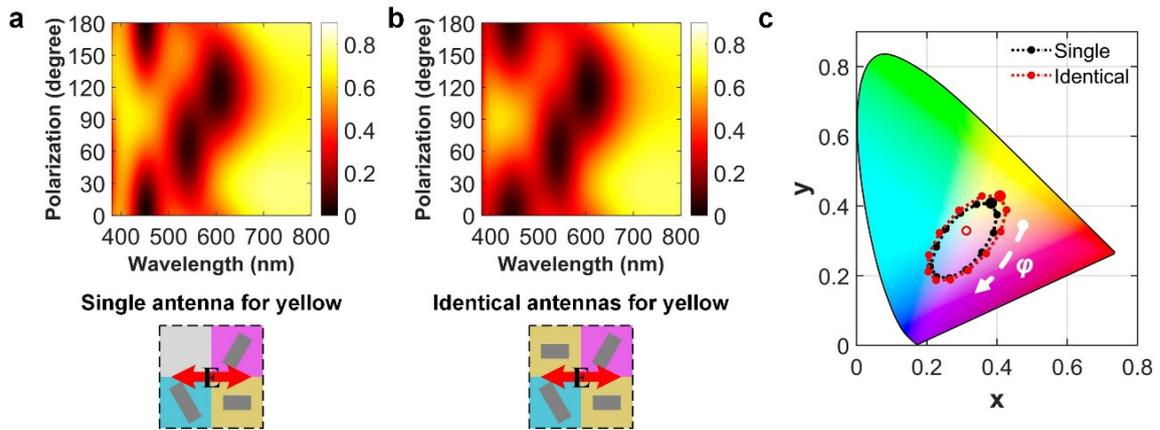

**Figure S1. Compensation of the low absorption cross-section of the nanoantenna for the yellow module**. Reflectance spectra and schematic of the full color tunable plasmonic pixels with (a) single nanoantenna of length 95 nm and (b) two identical nanoantennas of length 95 nm working for the yellow module. (c) Chromaticity coordinates of the full color tunable plasmonic pixels with different yellow modules. It is observed that the bandwidths of the reflection dips originating from the yellow modules are different for these two module layouts. The unit cell with two nanoantennas for the yellow module has broader reflection dips at the short wavelength, resulting in a higher saturation of the yellow color and a wider color gamut, as indicated by the red dotted line on the chromaticity diagram. Remarkably, to obtain a subtractive color with high color purity in the subtractive color model, a broad resonance bandwidth is highly desired.

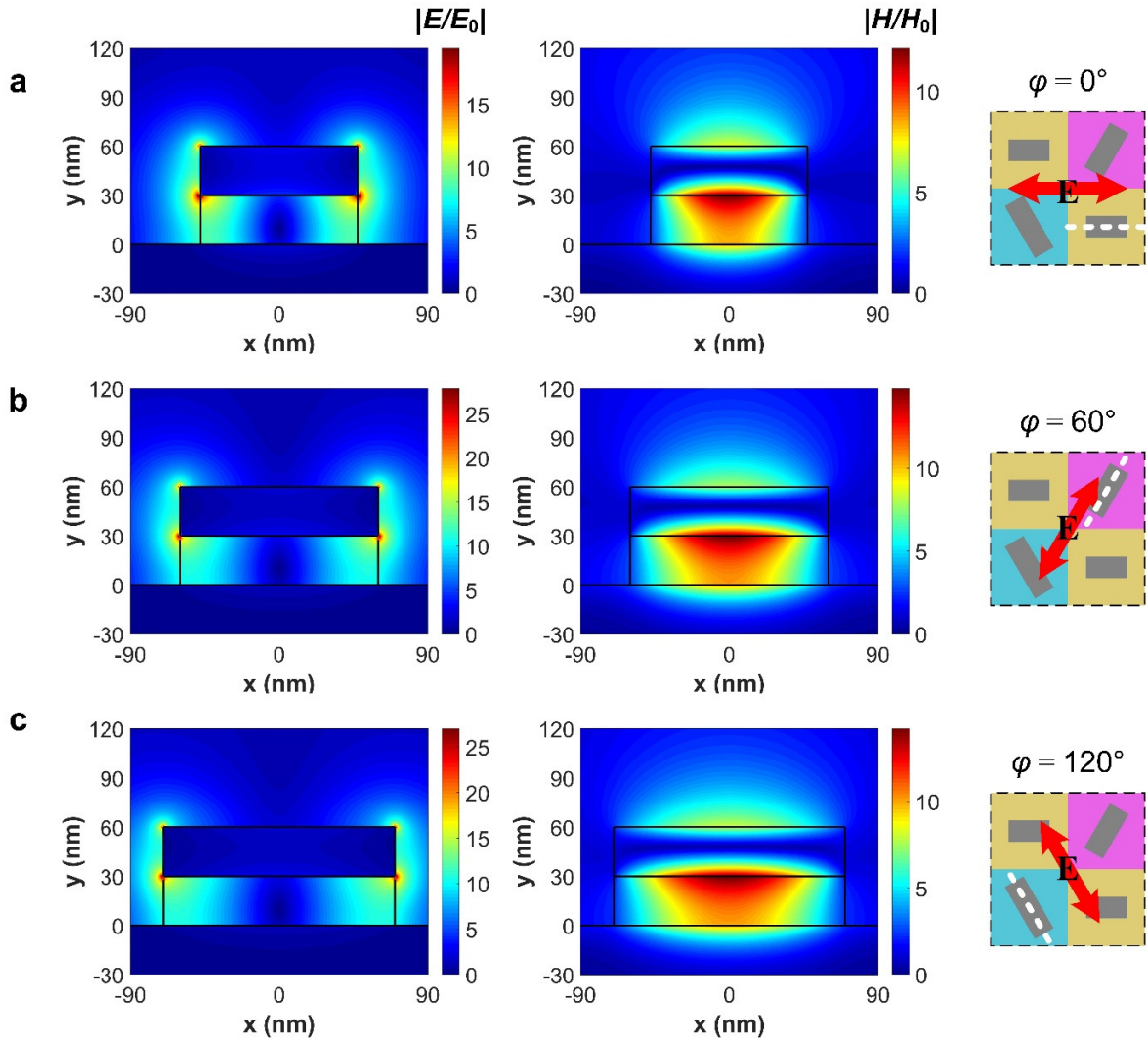

**Figure S2**. Spatial distributions of the simulated electrical field and magnetic field in the vertical cross-section along each color module for the polarization angles of (a) 0°, (b) 60° and (c) 120°, respectively. The electric field is distributed on both sides of the nanoantenna, especially near the corners of the metallic antenna. The magnetic field is confined in the spacer between the metallic antenna and the substrate. The electromagnetic field distributions indicate the excitation of gap surface plasmon for the incident light polarized along the corresponding nanoantennas. The insets show the configurations of the polarization direction of incident light and the corresponding vertical cross-section.

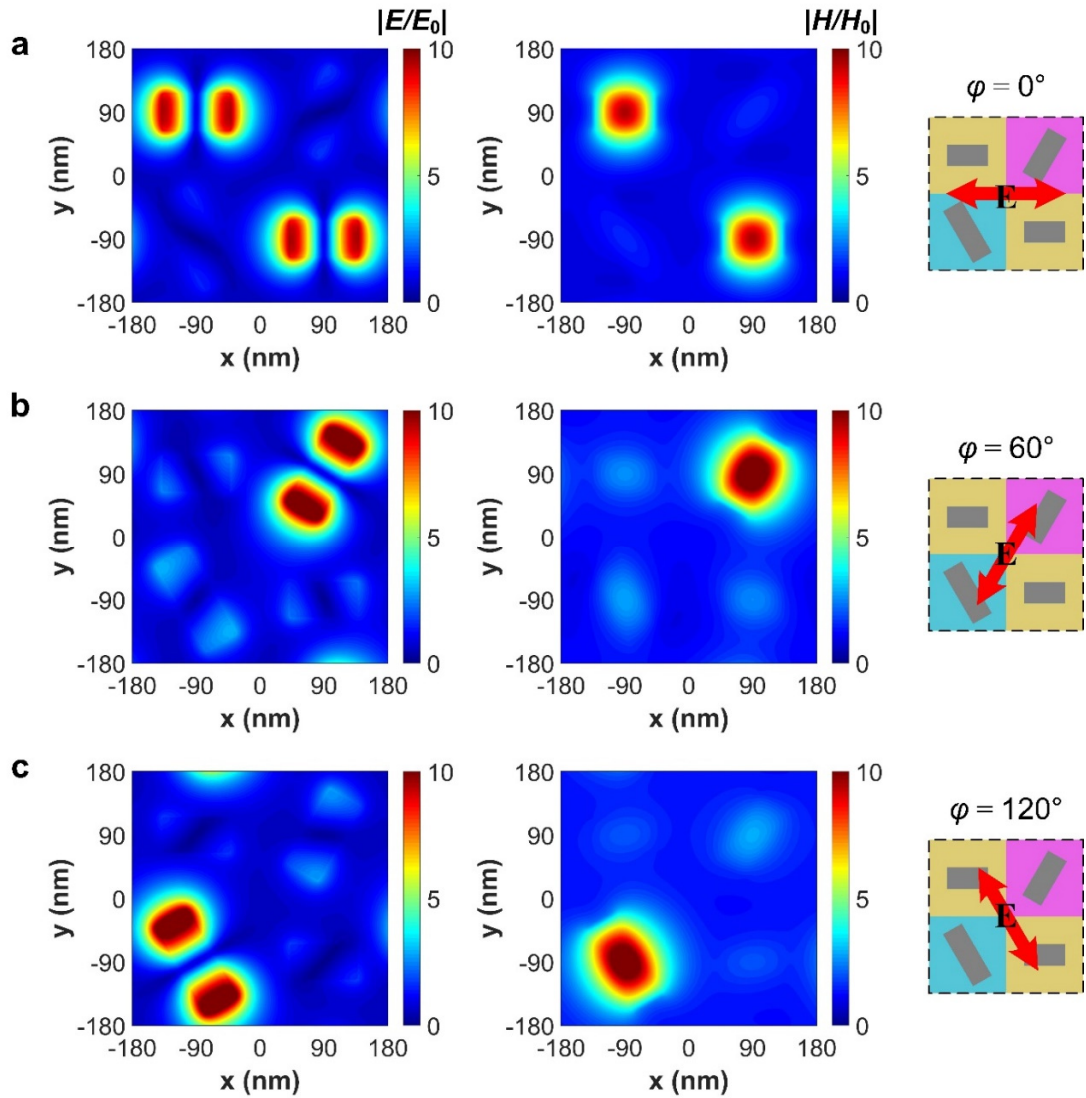

**Figure S3.** Spatial distributions of the simulated electrical field and magnetic field at the respective resonance wavelengths for the polarization angles of (a) 0°, (b) 60° and (c) 120°, respectively. The electromagnetic fields are strongly confined in the individual plasmonic nanoantennas for different polarizations with negligible coupling to adjacent nanoantennas. The insets show the configurations of the polarization-controlled full color tunable plasmonic pixels with the corresponding polarization direction of incident light. The upper limits of the colorbars are set at 10 for easy comparison.

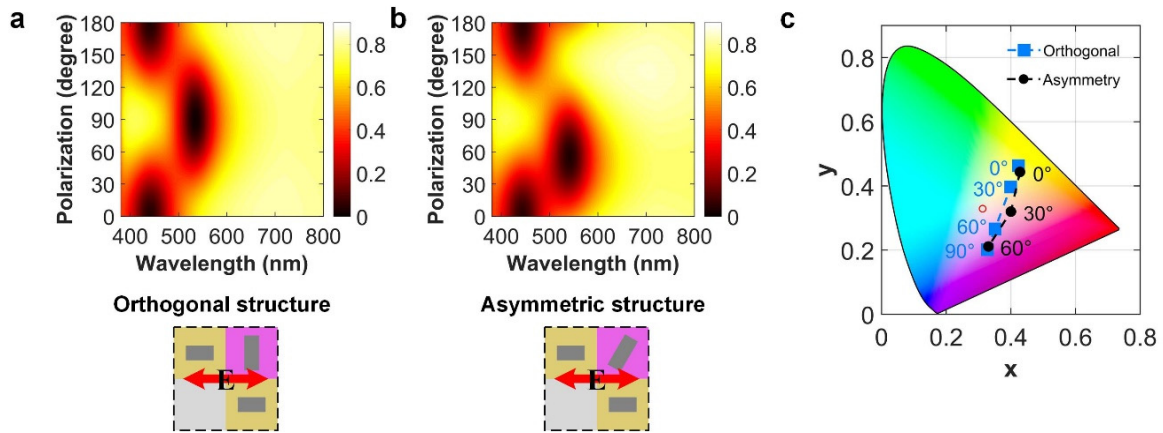

**Figure S4** Comparison of color tuning trajectory with increasing polarization angle. (a) Reflectance spectra of the orthogonal structure with the polarization angle ranging from 0° to 180°. The reflectance spectra are symmetric with respect to the polarization angle of 90° due to the two orthogonally placed color modules. (b) Reflectance spectra of the asymmetric nanostructure with the polarization angle varying from 0° to 180°. The two reflection dips are located at the polarization angles of 0° and 60° corresponding to the rotation angle of the two color modules. (c) Chromaticity coordinates of the orthogonal structure and asymmetric nanostructure with increasing polarization angle. As the incident polarization angle increases from 0° to 90°, the color tuning trajectory of the orthogonal structures are represented by a straight line between yellow and magenta on the chromaticity diagram. While the asymmetric nanostructures produce a curve path with increasing polarization angle from 0° to 60°, which is caused by the mixed spectra of the resonance dips for the non-orthogonal plasmonic nanoantennas.

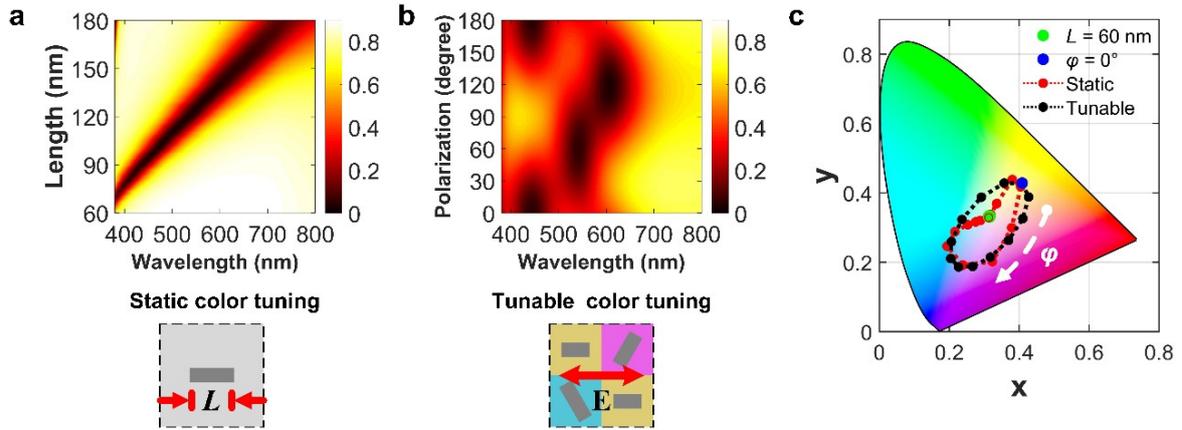

**Figure S5** Comparison of color gamut between static color change and polarization-controlled full color tuning. (a) Reflectance spectra of the nanorod arrays with lengths varying from 60 to 180 nm. The reflection dip of the static plasmonic pixel red-shifts with an increase in the length of the nanoantenna. (b) Reflectance spectra of the full color tunable plasmonic pixels as a function of polarization angle. The three reflection dips of the proposed plasmonic pixels are located at the polarization angles of 0°, 60°, and 120°, respectively. (c) Chromaticity coordinates of plasmonic pixels that undergo static color change and polarization-controlled full color tuning, respectively. The polarization-controlled full color tunable plasmonic pixels exhibit a wider color gamut than that of pixels that undergo static color change.

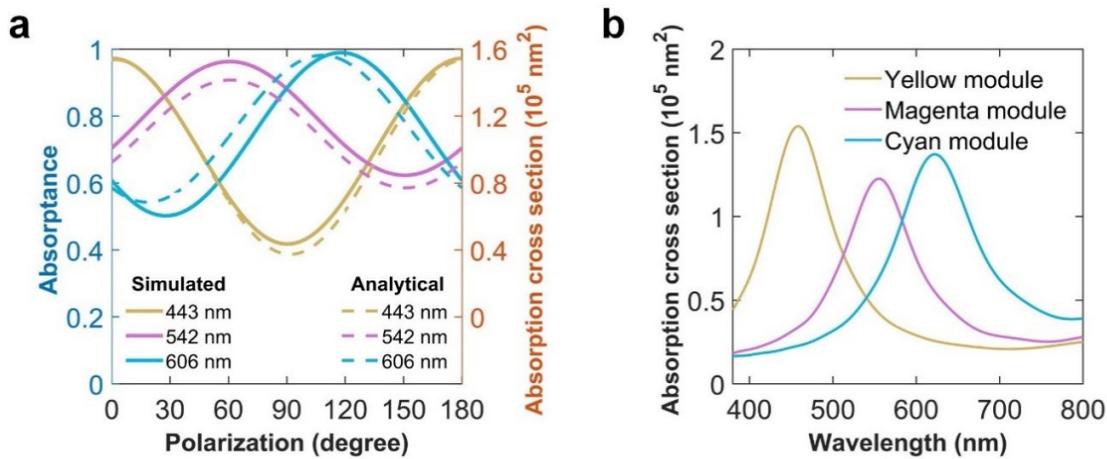

**Figure S6.** (a) Simulated absorptance versus polarization angle for three resonance wavelengths showing a cosine-squared shape (solid lines). The maximum absorptances are observed at the polarization angles of 0°, 60° and 120°, respectively, corresponding to the rotation angles of the color modules. The dash lines indicate the analytically calculated absorption cross-sections based on the equation at the resonance wavelengths, respectively. (b) Absorption cross-sections of various color modules for the polarization along the long axis. The relatively large absorption cross-section for the yellow module is due to the employment of two antennas.

The polarization dependent reflectance spectra of the FCTPPs are related to a linear combination of the polarization dependent absorption of three single color modules. As indicated by solid lines in Figure S6(a), the simulated absorptance as a function of polarization angle at three resonance wavelengths show a cosine-squared shape with maximums at the polarization angles of 0°, 60° and 120°, respectively. Moreover, the amplitude of absorptance oscillation at the wavelength of 443 nm is relatively large due to the large cross-section for the yellow module, which meets the purpose of using two antennas for the yellow module. Accordingly, a phenomenological expression can be written to estimate the polarization dependent spectra from the asymmetric nanostructures

$$1\text{-}R(\lambda,\varphi) \propto A_{\text{y}}(\lambda)\cos^2(\varphi) + A_{\text{m}}(\lambda)\cos^2(\varphi-60) + A_{\text{c}}(\lambda)\cos^2(\varphi-120)$$

where $\varphi$ is the polarization angle of the incident light, $\lambda$ is the free space wavelength of light, $A_{\text{y}}(\lambda)$, $A_{\text{m}}(\lambda)$ and $A_{\text{c}}(\lambda)$ are the maximum resonance absorption for incident light polarized along each color module in the asymmetric nanostructure, respectively. Here, the absorption cross-sections of three single color modules are used to represent the wavelength dependent constants $A_{\text{y}}(\lambda)$, $A_{\text{m}}(\lambda)$ and $A_{\text{c}}(\lambda)$, which are proportional to the resonance absorption. Moreover, two nanoantennas are used to calculate the absorption cross-section of the yellow modules, as shown in Figure S6(b). Although the absolute values are distinct, the shape of the simulated spectra (solid lines) and analytical absorption cross-sections (dash lines) are in good agreement, indicating that the polarization dependent properties of the asymmetric nanostructures can be well described by the analytical model.

The proposed phenomenological expression verifies two crucial issues. The first issue is that each nanoantenna can be deemed as independent color module due to the strong confinement property of the gap surface plasmons within the spacer, which is the fundamental of the modular design approach. The wavelength dependent constants Ay($\lambda$), Am($\lambda$) and Ac($\lambda$) in the phenomenological expression are represented by the absorption cross-sections of three color modules which are obtain from total-field scattered-field (TFSF) implementation using single color module without periodicity for the incident light polarized along the nanoantenna. The shape of the simulated spectra (solid lines) and the analytical calculations (dash lines) based on absorption cross-section are in good agreement, demonstrating that there is no significant plasmonic coupling between adjacent nanoantennas. One other issue is that the absorption response of the asymmetric plasmonic pixels to polarized light obeys the Malus' law, which is the basis of the polarization-controlled color tuning. When the incident light is polarized along each color module in the asymmetric nanostructure (the polarization angles of 0°, 60° and 120°), resonant dips in the reflection spectra occur, corresponding the excitation of gap-surface plasmons within the respective plasmonic nanoantennas. For arbitrary polarization, it gives a linear superposition of the three absorption cross-sections according to the projection of the polarization on the three different color modules. Thus, the phenomenological expression demonstrates how to analyze the polarization-dependent property of the asymmetric plasmonic pixels.

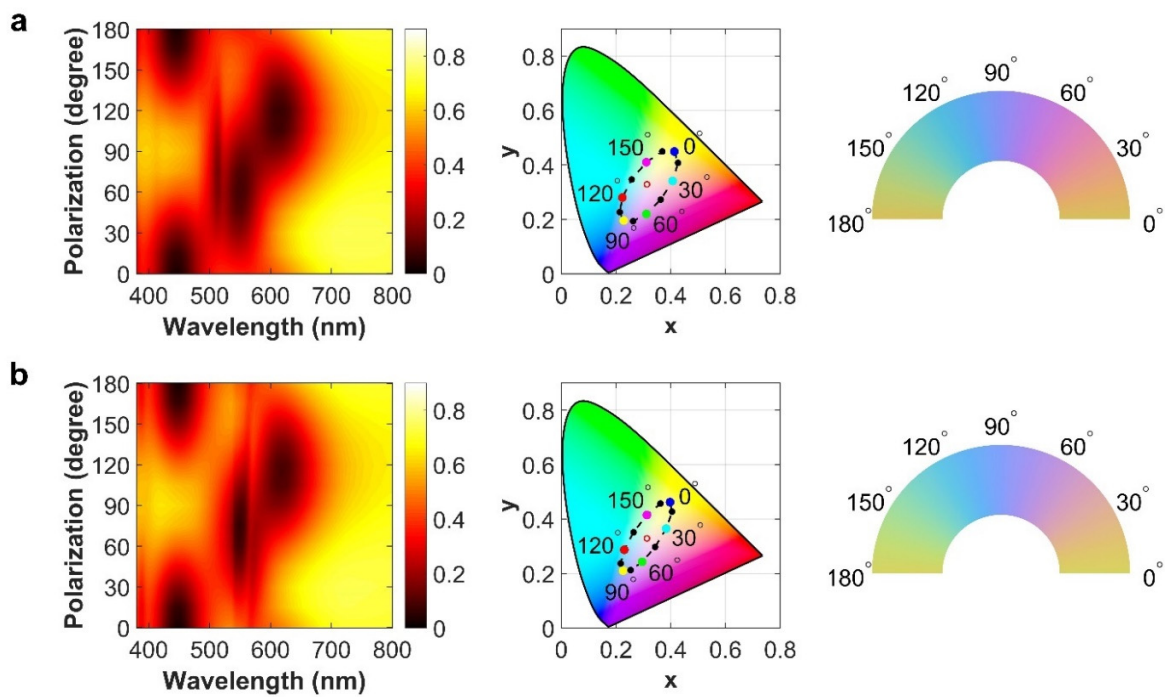

**Figure S7** The incident angular tolerance of the polarization-controlled full color tunable plasmonic pixels. (a) Simulated reflectance spectra and calculated colors of the plasmonic pixels as a function of polarization angle at incident angle of 20°. (b) Simulated reflectance spectra and calculated colors of the plasmonic pixels as a function of polarization angle at incident angle of 40°.

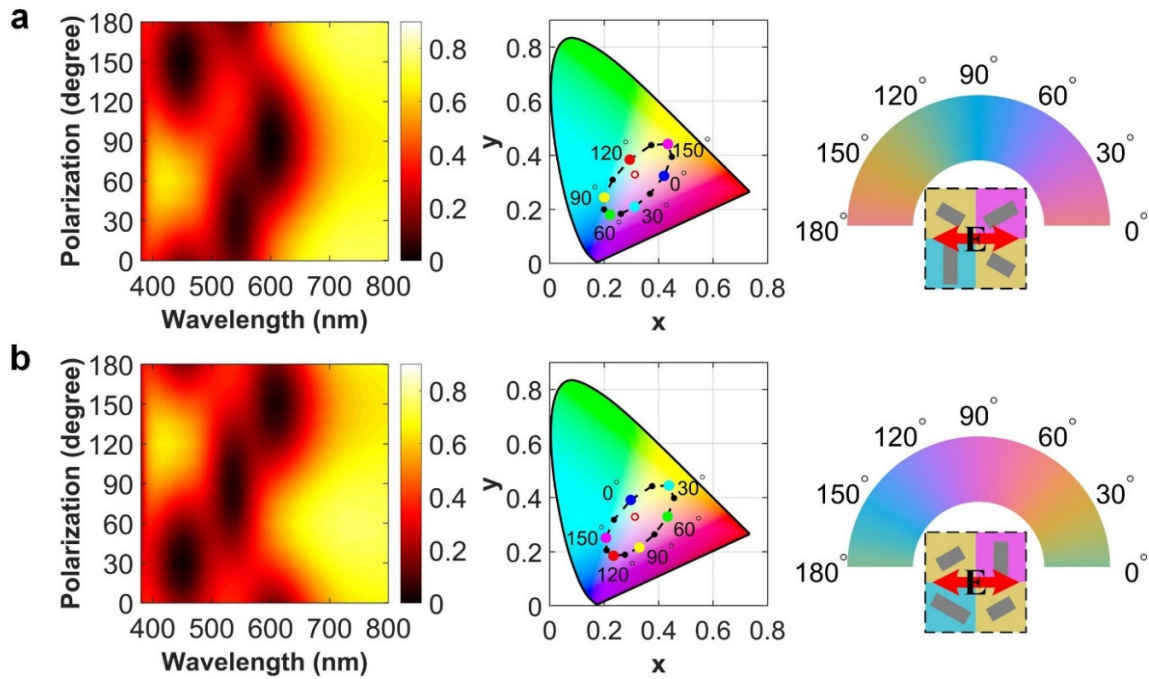

**Figure S8** (a) Polarization-controlled full color tuning with the initial color of red in clockwise order on the chromaticity diagram. The rotation angles for the yellow module, magenta module and cyan module are -30°, 30°, and 90°, respectively. (b) Polarization-controlled full color tuning with the initial color of green in clockwise order on the chromaticity diagram. The rotation angles for the yellow module, magenta module and cyan module are 30° and 90° and 150°, respectively. The corresponding module layouts are presented in the inset.

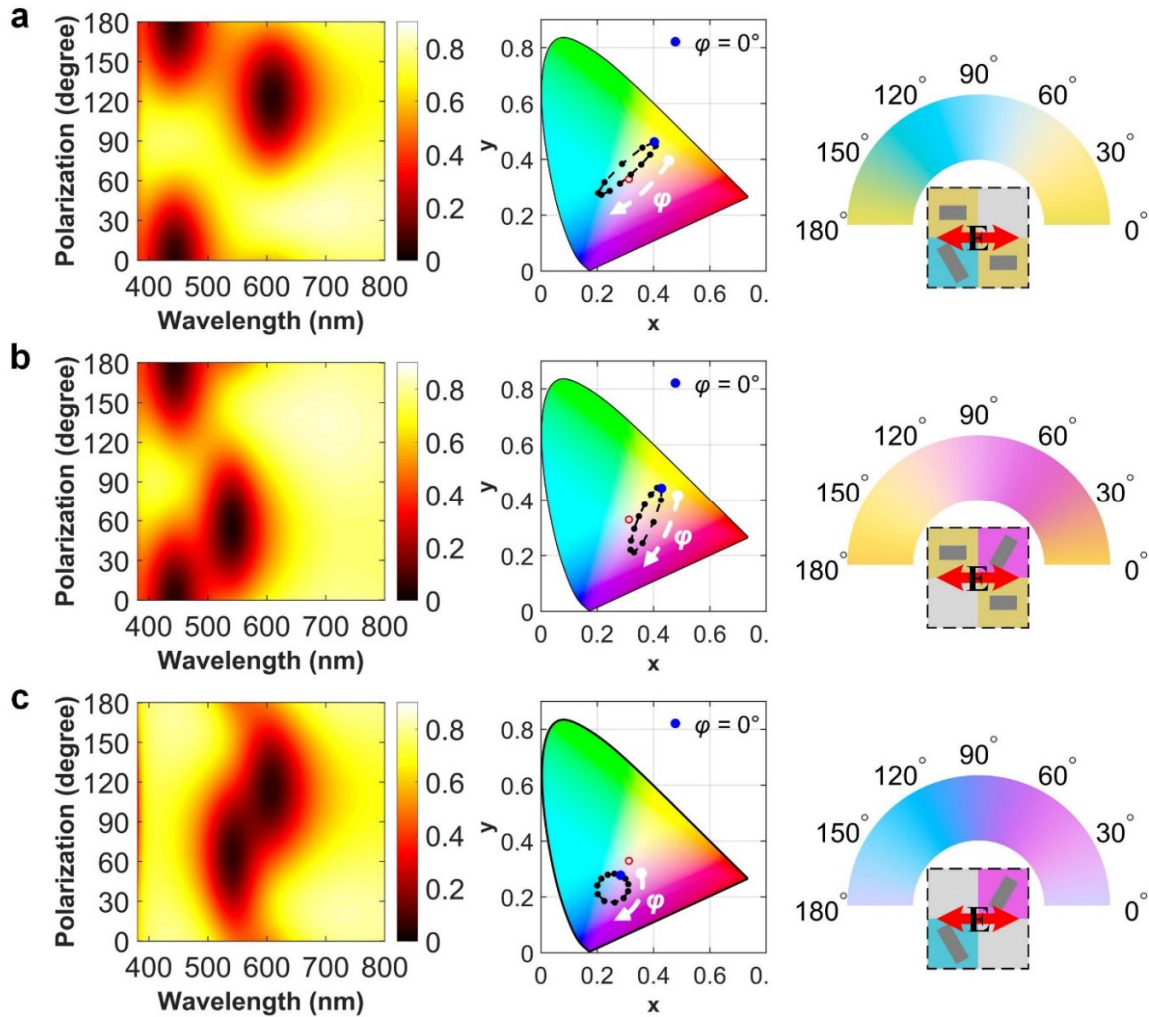

**Figure S9**. Polarization-controlled color tuning between two subtractive primary colors showing different color tuning ranges. Simulated reflectance spectra and evolutions of color tuning from (a) yellow to cyan, (b) yellow to magenta, and (c) magenta to cyan. The choices of component modules and the corresponding module layouts are presented in the inset. The color tuning trajectories for different module layouts demonstrate significantly different color tuning ranges with variations in the polarization angle of the incident light.

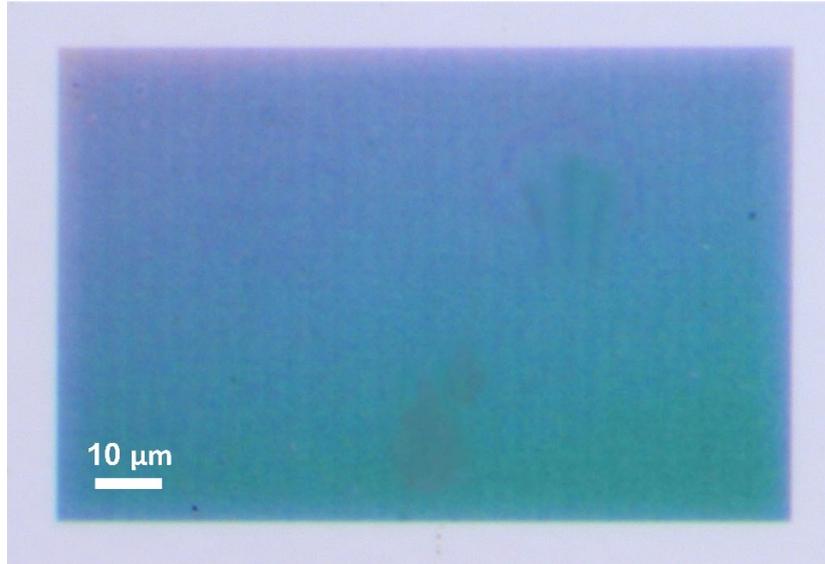

**Figure S10**. Optical micrographs of the microprint under illumination by natural light.

To further demonstrate the potential for information encryption in microprint encoded by various module layouts, we present the experimentally obtained optical micrographs of the microprint under illumination by natural light. The colors of the three patterns are almost the same as that of the background, because the polarization dependent plasmonic nanoantennas with different lengths and orientations in various module layouts are excited together under the unpolarized light. Upon a close scrutiny, there indeed exists a little color discrepancy for the three patterns and the background due to the fabrication imperfections such as rounded corners, orientation deviations and position deviations in the microprint. However, we believe the three patterns can be well concealed in the background if the resonance effects achieved with these layouts can be further optimized by improving the fabrication procedures to provide better accuracy and uniformity.

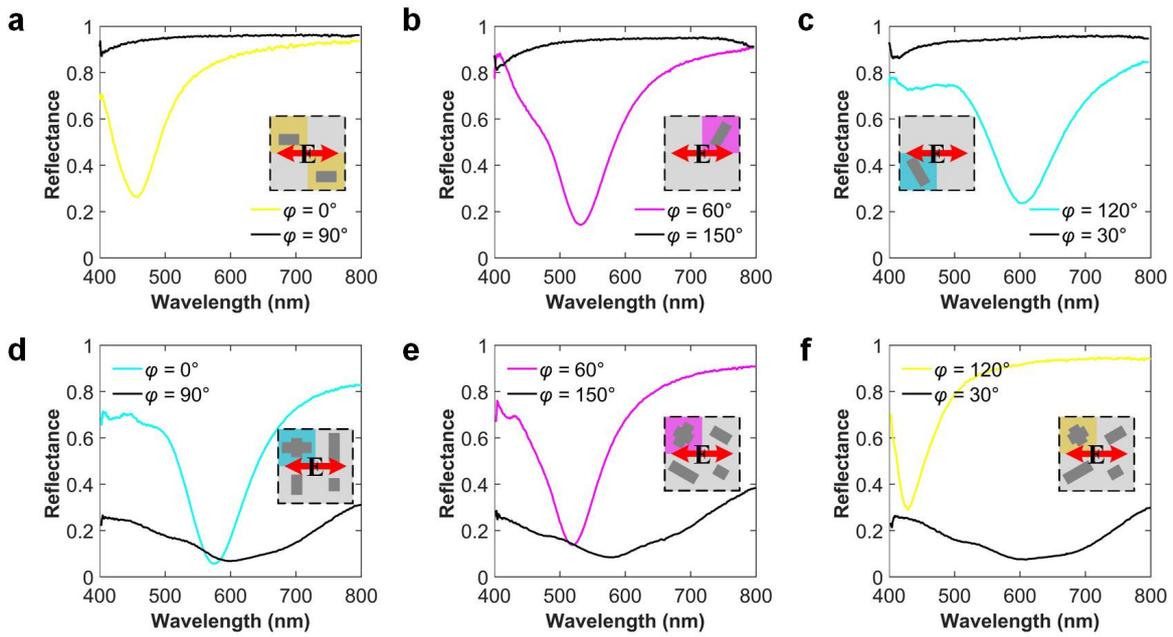

**Figure S11**. (a)-(c) Experimental reflectance spectra of the color switching between white and three subtractive primary colors of yellow, magenta and cyan at the specified polarization angles. (d)-(f) Experimental reflectance spectra of the color switching between black and three subtractive primary colors of cyan, magenta and yellow at the specified polarization angles.